\documentclass[12pt]{article}
\begin{document}
\begin{titlepage}
\begin{flushright}
hep-th/0410203\\
TIT/HEP-529\\
TU-734\\
October, 2004\\
\end{flushright}
\vspace{0.5cm}
\begin{center}
{\Large \bf Deformed Supersymmetry in Non(anti)commutative ${\cal N}=2$ 
Supersymmetric $U(1)$ Gauge Theory
}
\lineskip .75em
\vskip2.5cm
{\large Takeo Araki}${}^{1}$,\ \
{\large Katsushi Ito}${}^{2}$ \
and \ {\large Akihisa Ohtsuka}${}^{2}$
\vskip 2.5em
${}^{1}$ {\large\it Department of Physics\\
Tohoku University\\
Sendai, 980-8578, Japan}  \vskip 1.5em
${}^{2}${\large\it Department of Physics\\
Tokyo Institute of Technology\\
Tokyo, 152-8551, Japan}  \vskip 4.5em
\end{center}
\begin{abstract}
We study ${\cal N}=2$ supersymmetric $U(1)$ gauge theory in
non(anti)commutative ${\cal N}=2$  harmonic superspace with the 
chirality preserving non-singlet deformation parameter.
By solving the Wess-Zumino gauge preserving conditions for 
the analytic superfield,
we construct the deformed ${\cal N}=(1,0)$ supersymmetry
 transformation for component fields up to the first order in the
 deformation parameter.

\end{abstract}
\end{titlepage}

\baselineskip=0.7cm

Supersymmetric field theories in deformed superspace\cite{ncsuper} have been 
recently attracted much interest, partly motivated by studying superstring
effective field theories on the D-branes with graviphoton background
\cite{OoVa,BeSe,DeGrNi}.
Non(anti)commutative superspace is a deformed superspace with 
nonanticommutative Grassmann coordinates.
Field theories in non(anti)commutative ${\cal N}=1$ superspace
(${\cal N}=1/2$ superspace) is 
defined by the
fermionic version of the Moyal $*$-product. The deformed action can be
constructed in terms of superfields, whose procedure is the same as
field theories in noncommutative spacetime.
Compare to noncommutative field theories, the action usually contains a finite
number of deformed terms. The effects of non(anti)commutativity can 
be calculated explicitly.
There are a lot of works on field theories in deformed ${\cal N}=1$
superspace from  both perturbative and non-perturbative points of
view\cite{Se,ArItOh1,Pert,Inst}.

It is interesting to study the deformation of extended superspace 
\cite{KlPeTa}--
\cite{FeIvLeSoZu} 
because there is a variety of choices for the deformation.
In the case of the deformed ${\cal N}=2$ harmonic superspace, 
the deformation parameter can be decomposed into 
the singlet deformation part 
and the non-singlet part
with respect to $R$-symmetry group $SU(2)_R$ \cite{IvLeZu,FeSo}. 
In contrast to the case of ${\cal N}=1/2$ superspace, the deformed
action takes in general the form of infinite power series in the deformation
parameters, 
which is similar to noncommutative field theories.
In particular, 
${\cal N}=2$ supersymmetric gauge theory with the singlet deformation 
of ${\cal N}=2$ harmonic superspace 
has been recently studied in \cite{ArIt3} and \cite{FeIvLeSoZu}, 
where a field redefinition analogous to the Seiberg-Witten map \cite{SeWi}
in theories with space-space noncommutativity was found (see also \cite{Mi,SaWo}). 
The component action is fully determined in \cite{FeIvLeSoZu}. 

In the case of the non-singlet deformation parameterized by  $C$, 
we have studied the 
${\cal N}=2$ supersymmetric $U(1)$ gauge
theory in the non(anti)commutative 
${\cal N}=2$ harmonic superspace by using component formalism\cite{ArItOh2}.
Choosing the Wess-Zumino (WZ) gauge for the analytic superfield, 
we have written down the 
deformed action up to the first order in the deformation parameter.
We have shown that the commutative gauge transformation does not
preserve the WZ gauge due to the $*$-product 
and  one need to perform additional $C$-dependent
gauge transformation in order to recover the WZ gauge.
We have also made a field redefinition such that the component fields
transform canonically under the gauge transformation.

In this letter, we will study the chiral supersymmetry transformation 
(${\cal N}=(1,0)$ in the sense of \cite{IvLeZu}) of
the ${\cal N}=2$ supersymmetric $U(1)$ gauge theory in the 
chirality preserving 
non(anti)commutative ${\cal N}=2$ harmonic superspace.
Since two of the present authors discussed the exact gauge and
supersymmetry in the singlet deformation 
case \cite{ArIt3}, we will study the non-singlet deformation of the
superspace. 
We will determine the deformed supersymmetry up to the order $O(C)$
under which the $O(C)$ action in \cite{ArItOh2} is invariant.
The field redefinition given in \cite{ArItOh2}, 
which makes the deformed gauge transformation to be the same as the one 
in the ordinary abelian theory, is applied to the 
supersymmetry transformation. 
We will find that the component transformation laws are simplified 
by the redefinition. 

We begin with reviewing the non(anti)commutative ${\cal N}=2$
supersymmetric $U(1)$ gauge theory based on \cite{ArItOh2}.
Let ($x^{\mu},\theta_\alpha^i, \bar{\theta}_{\dot{\alpha}}^i$) be 
the coordinates of ${\cal N}=2$ (rigid) superspace.
Here $\mu=0,1,2,3$ are indices of spacetime with Euclidean signature.
$\alpha,\dot{\alpha}=1,2$ denote the spinor indices and $i=1,2$ labels
the  doublet of the $SU(2)_R$ $R$-symmetry.
We use the antisymmetric tensor $\varepsilon^{\alpha\beta}$ with
$\varepsilon^{12}=-\varepsilon_{12}=1$ for raising and lowering spinor
indices as in \cite{WeBa}
 but for $R$-symmetry indices we use $\epsilon^{ij}$ with
$\epsilon^{12}=-\epsilon_{21}=-1$.
In the Euclidean spacetime, $\theta^i_\alpha$ and
$\bar{\theta}^{i}_{\dot{\alpha}}$ are independent spinors.
The supersymmetry generators $Q^{i}_{\alpha}$,
$\bar{Q}_{\dot{\alpha}i}$ and the supercovariant
derivatives $D_{\alpha}^{i}$, $\bar{D}_{\dot{\alpha}i}$ are
defined by
\begin{eqnarray}
Q_{\alpha}^{i}&=&{\partial\over\partial\theta^{\alpha}_{i}}
-i(\sigma^{\mu})_{\alpha\dot{\alpha}}\bar{\theta}^{\dot{\alpha} i}
{\partial\over\partial x^{\mu}}, \quad
\bar{Q}_{\dot{\alpha} i}= -{\partial\over\partial\bar{\theta}^{\dot{\alpha} i}}
+i\theta^{\alpha}_{i}(\sigma^{\mu})_{\alpha\dot{\alpha}}
{\partial\over \partial x^{\mu}}
, 
\nonumber\\
D_{\alpha}^{i}&=&{\partial\over\partial\theta^{\alpha}_{i}}
+i(\sigma^{\mu})_{\alpha\dot{\alpha}}\bar{\theta}^{\dot{\alpha} i}
{\partial\over\partial x^{\mu}}, \quad
\bar{D}_{\dot{\alpha} i}= -{\partial\over\partial\bar{\theta}^{\dot{\alpha} i}}
-i\theta^{\alpha}_{i}(\sigma^{\mu})_{\alpha\dot{\alpha}}
{\partial\over \partial x^{\mu}}.
\end{eqnarray}
The ${\cal N}=2$ harmonic superspace \cite{GaIvOgSo} is introduced by adding 
the harmonic variables $u^{\pm}_i$ to the ${\cal N}=2$ superspace coordinates.
The variables $u^{\pm}_i$ form an $SU(2)$ matrix and satisfy the conditions
$ u^{+i}u^{-}_{i}=1$ and $\overline{u^{+i}}=u^{-}_{i}$.
The completeness condition for $u^{\pm}_i$ is given by
$u^{+}_{i}u^{-}_{j}-u^{+}_{j}u^{-}_{i}=\epsilon_{ij}$.
Using $u^{\pm}_{i}$, the $SU(2)_R$ indices can be projected into two
parts with $\pm 1$ $U(1)(\subset SU(2)_R)$ charges.
For example, we define the supercovariant derivatives 
$D^{\pm}_{\alpha}$ and $\bar{D}^{\pm}_{\alpha}$ by
$
 D^{\pm}_{\alpha}=u^{\pm}_{i}D^{i}_{\alpha}$, 
$\bar{D}^{\pm}_{\alpha}=u^{\pm}_{i}\bar{D}^{i}_{\alpha}
$.
$D^{i}_{\alpha}$ is solved by $D^{\pm}_{\alpha}$ such as
$D^{\pm}_{\alpha}=u^{+}_{i}D^{-}_{\alpha}-u^{-}_{i}D^{+}_{\alpha}$
with the help of the completeness condition.
In the harmonic superspace formalism, an important ingredient is an
analytic superfield rather than the ${\cal N}=2$ chiral superfield.
An analytic superfield $\Phi(x,\theta,\bar{\theta},u)$ is defined by
$D^{+}_{\alpha}\Phi=\bar{D}^{+}_{\dot{\alpha}}\Phi=0$.
It is convenient to write this analytic superfield in terms of analytic
basis:
$
 x_{A}^{\mu}= 
x^{\mu}-i (\theta^i \sigma^\mu \bar{\theta}^j +\theta^j
\sigma^\mu\bar{\theta}^i)
u^{+}_{i}u^{-}_{j}$,
$\theta^{\pm}_{\alpha}= u^{\pm }_{i}\theta^{i}_{\alpha}$ and 
$\bar{\theta}^{\pm}_{\dot{\alpha}}= u^{\pm}_{i}\bar{\theta}^{i}_{\dot{\alpha}}$.
In the analytic basis, 
an analytic superfield $\Phi$ is functions
of $(x^{\mu}_{A},\theta^{+},\bar{\theta}^{+},u)$: 
$
 \Phi=\Phi(x^{\mu}_{A},\theta^{+},\bar{\theta}^{+},u)
$. 
We now introduce the nonanticommutativity in the ${\cal N}=2$ harmonic
superspace by using the $*$-product:
\begin{equation}
 f*g(\theta)=f(\theta)\exp(P) g(\theta),\quad
P=-{1\over2}
\overleftarrow{Q^{i}_{\alpha}}
C^{\alpha\beta}_{ij}\overrightarrow{Q^{j}_{\beta}}, 
\label{eq:moyal2}
\end{equation}
where $C^{\alpha\beta}_{ij}$ is some constants. 
With this $*$-product, we have following (anti)commutation relations: 
\begin{equation}
 \{ \theta^{\alpha}_{i}, \theta^{\beta}_{j}\}_{*}=C^{\alpha\beta}_{ij}
, \quad 
 [x^{\mu}_{L}, x^{\nu}_{L}]_{*}=[x^{\mu}_{L}, \theta^{\alpha}_{i}]_{*}=
[x^{\mu}_{L},\bar{\theta}^{\dot{\alpha} i}]_{*}=0, \quad 
\{ \bar{\theta}^{\dot{\alpha} i}, \bar{\theta}^{\dot{\beta} j}\}_{*}=
\{ \bar{\theta}^{\dot{\alpha} i}, \theta^{\alpha}_{j}\}_{*}=0
, 
\end{equation}
where $x_{L}^{\mu}\equiv x^\mu+i\theta_i \sigma^\mu \bar{\theta}^i$. 
The deformation parameter $C^{\alpha\beta}_{ij}$ is symmetric
under the exchange of pairs of indices $(\alpha i)$,$(\beta j)$:
 $C^{\alpha\beta}_{ij}=C^{\beta\alpha}_{ji}$.
We decompose the nonanticommutative parameter $C^{\alpha\beta}_{ij}$
into the symmetric and antisymmetric parts with respect to the $SU(2)$ indices,
such as
\begin{equation}
C^{\alpha\beta}_{ij}=C^{\alpha\beta}_{(ij)}
+{1\over4}\epsilon_{ij}\varepsilon^{\alpha\beta}C_{s}.
\end{equation}
Here we denote $A_{(i_1\cdots i_n)}$ by the symmetrized sum of
$A_{i_1\cdots i_n}$ over indices $i_1,\cdots, i_n$.
$C^{\alpha\beta}_{ij}$ with zero $C^{\alpha\beta}_{(ij)}$ 
corresponds to the singlet deformation \cite{IvLeZu,FeSo}.
For superfields $A$ and $B$, the $*$-product takes the form
\begin{equation}
 A*B=AB+APB+{1\over2}AP^2B+{1\over6}AP^3B+{1\over24}AP^4B,\quad
P^5=0.
\label{eq:star2}
\end{equation}
Since $P$ commutes with the supercovariant derivatives 
$D$, the chiral structure is preserved by this deformation.
In the analytic basis, one can compute the $*$-product by using
$Q^{i}_{\alpha}=u^{+i}Q^{-}_{\alpha}-u^{-i}Q^{+}_{\alpha}$.
For example we have
\begin{equation}
\{ \theta^{\eta},\theta^{\eta'} \}_{*}=C^{\eta \eta' \alpha\beta},
\quad 
 \mbox{[}x^{\mu}_{A}, x^{\nu}_{A}\mbox{]}_{*}= 4 C^{--\mu\nu}
(\bar{\theta}^{+})^2
, \quad 
\mbox{[}x^{\mu}_{A},
\theta^{\eta}_{\alpha} \mbox{]}_{*}=
-2i C^{-\eta \beta\alpha}(\sigma^{\mu}\bar{\theta}^{+})_{\beta}
, 
\end{equation}
where $\eta,\eta'=\pm$, 
$C^{\eta\eta'\mu\nu}=u^{\eta i}u^{\eta' j}C^{\mu\nu}_{ij}$,
$C^{\mu\nu}_{ij}\equiv C^{\alpha\beta}_{ij}\sigma^{\mu\nu}_{\alpha}
{}^{\gamma} \varepsilon_{\beta\gamma}$ and
$\sigma^{\mu\nu}={1\over4}(\sigma^{\mu}\bar{\sigma}^{\nu}
-\sigma^{\nu}\bar{\sigma}^{\mu})$. 
Since 
we will consider the non-singlet deformation, we put $C_s=0$ 
in the following.

We now construct the action of ${\cal N}=2$ supersymmetric $U(1)$
gauge theory in this non(anti)commutative superspace.
We introduce an analytic superfield $V^{++}(\zeta,u)$ 
with $\zeta=(x_A^{\mu},\theta^+, \bar{\theta}^+)$ by covariantizing
the harmonic derivative 
$D^{++}
= 
        u^{+i}{\partial\over \partial u^{-i}}
        - 2i \theta^+ \sigma^\mu \bar{\theta}^+ 
                {\partial\over \partial x_A^\mu } 
        + \theta ^{+\alpha} {\partial \over \partial \theta^{-\alpha}} 
        + \bar{\theta}^{+\dot{\alpha}} 
                {\partial \over \partial \bar{\theta}^{-\dot{\alpha}}}
\rightarrow \nabla^{++}=D^{++}+i V^{++}$.
Generalizing the construction in \cite{Zu,BuSa}, the action is given by
\begin{equation}
S_{*}
= 
\frac12 \sum_{n=2}^{\infty} \int d^4xd^8\theta du_1\dots du_n {(-i)^n \over n}
{ V^{++}(1) * \cdots * V^{++}(n) 
\over (u_1^+ u_2^+)\cdots (u_{n}^+ u_1^+) }
. 
\label{eq:StarDeformedAction:Gen}
\end{equation} 
where $V^{++}(i)=V^{++}(\zeta_i, u_i)$,
$\zeta_{i}=(x_{A},\theta^{+}_{i},\bar{\theta}^{+}_{i})$ and 
$d^8\theta=d^4 \theta^+ d^4\theta^- $ with
$d^4\theta^{\pm}=d^2\theta^{\pm}d^2\bar{\theta}^{\pm}$.
The harmonic integral is defined by the rules:
(i)
$
 \int du f(u)=0
$
for $f(u)$ with non-zero $U(1)$ charge.
 (ii)
$
 \int du 1=1.
$
(iii)
$
 \int du u^{+}_{(i_1}\cdots u^{+}_{i_n} u^{-}_{j_1}\cdots u^{-}_{j_n)}=0,
\ (n\geq 1).
$
The action (\ref{eq:StarDeformedAction:Gen}) is invariant under the 
gauge transformation
\begin{equation}
 \delta_\Lambda^{*} V^{++}=-D^{++}\Lambda+i [ \Lambda, V^{++}]_{*}
, 
\label{eq:gauge2}
\end{equation}
with an analytic superfield $\Lambda$.
The generic superfield $V^{++}(\zeta,u)$ includes infinitely many
auxiliary fields. 
Most of these fields are gauged away except the lowest component fields 
in the harmonic expansion. 
One can take the WZ gauge
\begin{eqnarray}
 V^{++}_{WZ}(x_{A},\theta^{+},\bar{\theta}^{+},u)
&=& 
-i\sqrt{2}(\theta^{+})^2 \bar{\phi}(x_{A})
+i\sqrt{2}(\bar{\theta}^{+})^2 \phi(x_{A})
-2i \theta^{+}\sigma^{\mu}\bar{\theta}^{+}A_{\mu}(x_{A})\nonumber\\
&&+4(\bar{\theta}^{+})^2\theta^{+}\psi^{i}(x_{A}) u^{-}_{i}
-4(\theta^{+})^2\bar{\theta}^{+}\bar{\psi}^{i}(x_{A})u^{-}_{i}\nonumber\\
&&
+3(\theta^{+})^2(\bar{\theta}^{+})^2 D^{ij}(x_{A})u^{-}_{i} u^{-}_{j},
\label{eq:wz1}
\end{eqnarray}
which is convenient to study the theory in the component formalism. 

The component action $S_*$ in the WZ gauge can be expanded in a power series
of the deformation parameter $C$. 
In \cite{ArItOh2}, we have computed the $O(C)$ action explicitly.
The quadratic part $S_{*,2}$ in $S_{*}$ is the same as the commutative
one:
\begin{equation}
S_{*,2}=\int d^4 x\left\{
-{1\over 4}F_{\mu\nu}F^{\mu\nu}-{1\over4}F_{\mu\nu}\tilde{F}^{\mu\nu}
+\phi \partial^2 \bar{\phi}
-i\psi^{i}\sigma^{\mu}\partial_{\mu}\bar{\psi}_{i}
+{1\over4}D^{ij}D_{ij}
\right\}.
\label{eq:abelianaction}
\end{equation}
The cubic part $S_{*,3}$ in $S_{*}$ is of order $O(C)$ and given by
\begin{eqnarray}
S_{*,3}
&=&{}
        \int d^4x \left[
- { 2 \sqrt{2} \over 3} i C_{(ij)}^{\alpha\beta}
        \psi^i_\alpha ( \sigma^\nu \partial_\nu \bar{\psi}^j)_\beta 
        \bar{\phi}
- 2\sqrt{2} i C_{(ij)}^{\alpha\beta}
        \psi^i_\alpha (\sigma^\nu \bar{\psi}^j)_\beta 
        \partial_\nu \bar{\phi}
        \right. \nonumber\\ 
&&{}
+{2\over 3} i C_{(ij)}^{\alpha\beta} A_\mu
        (\sigma^\mu \bar{\psi}^i)_\alpha (\sigma^\nu \partial_\nu \bar{\psi}^j)_\beta
- i C_{(ij)}^{\mu\nu} \bar{\psi}^i \bar{\psi}^j  F_{\mu\nu}
\nonumber\\ 
&&\left. {} 
+ \sqrt{2} C_{(ij)}^{\mu\nu} D^{ij} A_\mu \partial_\nu \bar{\phi}
+ {1\over \sqrt{2}} C_{(ij)}^{\mu\nu} D^{ij} F_{\mu\nu} \bar{\phi} 
\right] 
. 
\label{eq:thirdorderLagrangian}
\end{eqnarray}
Note that here we have already dropped the $C_s$ dependent terms. 
We will refer 
\begin{equation} 
S_{*,2}+ S_{*,3} 
\label{eq:O(C)action} 
\end{equation} 
as the $O(C)$ action. 

In the commutative case, the gauge parameter $\Lambda=\lambda(x_A)$
preserves the WZ gauge and gives rise to the gauge transformation for
component fields.
In the non(anti)commutative case, however, 
the gauge transformation (\ref{eq:gauge2}) with the same gauge
parameter does not preserve the WZ gauge because of the $C$-dependent
terms arising from the commutator. 
In order to preserve the WZ gauge, one must include the $C$-dependent 
terms. The gauge parameter
is shown to take the form 
\begin{eqnarray}
\lambda_C (\zeta, u)
&=&
        \lambda(x_A)
        + \theta^{+}\!\sigma^\mu \bar{\theta}^{+}
                \lambda_\mu^{(-2)} (x_A, u; C)
        + (\bar{\theta}^{+})^2 
                \lambda^{(-2)} (x_A, u; C)
                \nonumber\\
&&{} 
        + (\bar{\theta}^{+})^2 \theta^{+}{}^{\alpha} 
                \lambda_{\alpha}^{(-3)} (x_A, u; C)
        + (\theta^{+})^2 (\bar{\theta}^{+})^2 
                \lambda^{(-4)} (x_A, u; C)
, 
\label{eq:gaugeparam1}
\end{eqnarray}
which has been determined by solving the 
WZ gauge preserving conditions expanded  in harmonic modes
\cite{ArItOh2}. 
The gauge transformation is also fully determined, which reads 
\begin{eqnarray}
\delta^{*}_{\lambda_C} A_\mu 
&=& 
        - \partial_\mu \lambda 
        + O(C^2),
\nonumber\\
\delta^{*}_{\lambda_C} \phi 
&=&
        O(C^2),
\nonumber\\
\delta^{*}_{\lambda_C} \psi_{\alpha i} 
&=& 
        \frac23 (\varepsilon C_{(ij)} \sigma^\mu \bar{\psi}^j )_\alpha 
        \, \partial_{\mu} \lambda 
        +O(C^2) ,
\nonumber\\
\delta^{*}_{\lambda_C} D_{ij}
&=&
        2 \sqrt{2} C_{(ij)}^{\mu\nu} \partial_\mu \lambda \partial_\nu \bar{\phi} 
        +O(C^2) \nonumber\\
\delta_{\lambda_C}^{*} (\mbox{others}) 
&=& 
        0
        . 
\label{eq:gaugetr1}
\end{eqnarray}
The $O(C)$ action is 
invariant under the $O(C)$ gauge transformation (\ref{eq:gaugetr1}).

These gauge transformations are not canonical.
But if we redefine the component fields such as
\begin{eqnarray}
\hat{A}_\mu 
&=& 
        A_\mu 
        + O(C^2) 
        ,\nonumber\\ 
\hat{\phi} 
&=& 
        \phi 
        + O(C^2)
        , 
\quad \hat{\bar{\phi}}=\bar{\phi},
\nonumber\\ 
\hat{\psi}_{\alpha i} 
&=& 
        \psi_{\alpha i} 
        +\frac23 (\varepsilon C_{(ij)} \sigma^\mu \bar{\psi}^j )_\alpha A_{\mu} 
        +O(C^2) 
        , \quad
\hat{\bar{\psi}}{}^{\dot{\alpha}}=\bar{\psi}^{\dot{\alpha}}
\nonumber\\
\hat{D}_{ij} 
&=&  
        D_{ij} 
        + 2 \sqrt{2} C_{(ij)}^{\mu\nu} A_\mu \partial_\nu \bar{\phi} 
        +O(C^2),
\label{eq:orderCgaugetr}
\end{eqnarray}
the newly defined fields are shown to transform canonically: $
\delta_{\lambda_C}^* \hat{A}_{\mu}=-\partial_{\mu}\lambda$, 
$\delta_{\lambda_C}^* (\mbox{others})=0$.
In terms of redefined fields, the $O(C)$ action can be written as
\begin{eqnarray}
S_{*,2} + S_{*,3} 
&=& 
        \int d^4x \left[
        - {1\over 4} 
                \hat{F}_{\mu\nu} ( \hat{F}^{\mu\nu} + \tilde{\hat{F}}{}^{\mu\nu} ) 
        + \hat{\phi} \partial^2 \hat{\bar{\phi}}
        -i \hat{\psi}^i \sigma^\mu \partial_\mu \hat{\bar{\psi}}_i
        + \frac14 
                 \hat{D}^{ij} \hat{D}_{ij}
                \right. \nonumber\\
&&{}
        - 2\sqrt{2} i C_{(ij)}^{\alpha\beta}
                \hat{\psi}^i_\alpha (\sigma^\mu \hat{\bar{\psi}}{}^j)_\beta 
                \partial_\mu \hat{\bar{\phi}}
        - { 2 \sqrt{2} \over 3} i C_{(ij)}^{\alpha\beta}
        \hat{\psi}^i_\alpha ( \sigma^\mu \partial_\mu 
\hat{\bar{\psi}}{}^j)_\beta 
                \hat{\bar{\phi}}
        \nonumber\\
&&\left.{}
        - i C_{(ij)}^{\mu\nu} \hat{\bar{\psi}}{}^i \hat{\bar{\psi}}{}^j 
\hat{F}_{\mu\nu}
        + {1\over \sqrt{2}} C_{(ij)}^{\mu\nu} \hat{D}^{ij} \hat{F}_{\mu\nu}
\hat{\bar{\phi} }
        + O(C^2)
        \right] 
,
\end{eqnarray}
where $\hat{F}_{\mu\nu}=\partial_{\mu}\hat{A}_{\nu}
-\partial_{\nu}\hat{A}_{\mu}$.

Now we study the supersymmetry transformation 
that is generated by the chiral part of the supersymmetry generators: 
the ${\cal N}=(1,0)$ supersymmetry generated by $Q_\alpha^i$. 
The deformed supersymmetry transformation of the gauge multiplet, 
\begin{eqnarray}
\delta^{*}_{\xi} V^{++}_{WZ}
&=&  
        -i\sqrt{2}(\theta^{+})^2 \delta^{*}_{\xi} \bar{\phi}(x_{A})
        +i\sqrt{2}(\bar{\theta}^{+})^2 \delta^{*}_{\xi} \phi(x_{A})
        -2i \theta^{+}\sigma^{\mu}\bar{\theta}^{+} 
                \delta^{*}_{\xi} A_{\mu}(x_{A})
                \nonumber\\
&&{}
        +4(\bar{\theta}^{+})^2\theta^{+} \delta^{*}_{\xi} \psi^{i}(x_{A}) u^{-}_{i}
        -4(\theta^{+})^2\bar{\theta}^{+} 
                \delta^{*}_{\xi} \bar{\psi}^{i}(x_{A})u^{-}_{i}
                \nonumber\\
&&{}
        +3(\theta^{+})^2(\bar{\theta}^{+})^2 
                \delta^{*}_{\xi} D^{ij}(x_{A})u^{-}_{i} u^{-}_{j}
, 
\end{eqnarray}
is given by 
\begin{equation}
\delta^{*}_{\xi} V^{++}_{WZ} 
= 
        \tilde{\delta}_\xi V^{++}_{WZ} 
        + \delta^{*}_\Lambda V^{++}_{WZ} 
, 
\label{eq:DeformedSUSY}
\end{equation}
where 
\begin{equation} 
\tilde{\delta}_{\xi}V^{++}_{WZ}
= 
        \left(
        -\xi^{+\alpha}Q^{-}_{\alpha}+\xi^{-\alpha}Q^{+}_{\alpha}
        \right) 
        V^{++}_{WZ}
\end{equation} 
and $\delta^{*}_{\Lambda} V^{++}_{WZ}$ is 
a deformed gauge transformation of $V^{++}_{WZ}$ 
with an appropriate analytic gauge parameter 
$\Lambda(\zeta,u)$ 
to retain the WZ gauge: 
\begin{equation}
\delta^{*}_{\Lambda} V^{++}_{WZ} (\zeta, u)
= 
        -D^{++} \Lambda (\zeta,u) 
        + i [ \Lambda, V^{++}_{WZ} ] _{*}(\zeta, u)
        . 
\end{equation}
We will denote the analytic gauge parameter as 
\begin{eqnarray}
\Lambda(\zeta, u) 
&=& 
        \lambda^{(0,0)} (x_A,u)
+ \bar{\theta}^{+}_{\dot{\alpha}} \lambda^{(0,1)}{}^{\dot{\alpha}} (x_A,u)
        + \theta^{+ \alpha} \lambda^{(1,0)}_{\alpha} (x_A,u)
        + (\bar{\theta}^{+})^2 \lambda^{(0,2)} (x_A, u)
        \nonumber\\
&&{} 
        + (\theta^{+})^2 \lambda^{(2,0)} (x_A, u)
        + \theta^{+}\!\sigma^\mu \bar{\theta}^{+} 
                \lambda_\mu^{(1,1)} (x_A, u)
        + (\bar{\theta}^{+})^2 \theta^{+}{}^{\alpha} 
                \lambda^{(1,2)}_{\alpha} (x_A, u)
        \nonumber\\
&&{} 
        + (\theta^{+})^2 \bar{\theta}^{+}_{\dot{\alpha}} 
                \lambda^{(2,1)}{}^{\dot{\alpha}} (x_A, u)
        + (\theta^{+})^2 (\bar{\theta}^{+})^2 
                \lambda^{(2,2)} (x_A, u) 
, 
\end{eqnarray} 
where $\lambda^{(n,m)} (x_A, u)$ 
is the $(\theta^{+})^{n} (\bar{\theta}^{+})^{m}$-component.

{}From eq. (\ref{eq:DeformedSUSY}), 
the equations to determine the deformed supersymmetry transformation laws are 
obtained as follows: 
\begin{eqnarray}
0 
&=& 
        2 i (\xi^{+} \sigma^\mu)_{\dot{\beta}} 
                        \varepsilon^{\dot{\beta}\dot{\alpha}} A_\mu 
        - \partial^{++} \lambda^{(0,1)}{}^{\dot{\alpha}} 
        - 2 \lambda^{(1,0)}{}^{\alpha} 
                        (\varepsilon C^{++} \sigma^\mu )_{\alpha{\dot{\beta}}}
                        \varepsilon^{{\dot{\beta}}{\dot{\alpha}}} 
                        A_\mu 
        , 
        \label{eq:EqsToDetermineDeformedSUSY(0,1)}
        \\
0 
&=& 
        - 2 \sqrt{2} i \xi^{+}_{\alpha} \bar{\phi} 
        - \partial^{++} \lambda^{(1,0)}_\alpha 
        - 2 \sqrt{2} (\varepsilon C^{++} \lambda^{(1,0)} )_\alpha \bar{\phi} 
        , 
        \label{eq:EqsToDetermineDeformedSUSY(1,0)}
        \\
\sqrt{2} i 
\delta^{*}_{\xi} \phi 
&=& 
        4 \xi^{+} \psi^i u_i^{-}
        + 4 i \lambda^{(1,0)}{}^{\alpha} (\varepsilon C^{++} \psi^i )_\alpha u_i^{-} 
        \nonumber\\
&&{} 
        - \partial^{++} \lambda^{(0,2)} 
        - C^{++}{}^{\alpha\beta} 
                        (\sigma^\nu \bar{\sigma}^\mu \varepsilon )_{\alpha\beta} 
                        \lambda_\mu^{(1,1)} A_\nu 
        , 
        \label{eq:EqsToDetermineDeformedSUSY(0,2)}
        \\
- \sqrt{2} i 
\delta^{*}_{\xi} \bar{\phi}
&=& 
        0 
        , 
        \label{eq:EqsToDetermineDeformedSUSY(2,0)}
        \\
- 2 i 
\delta^{*}_{\xi} A_\mu 
&=& 
        4 \xi^{+} \sigma_\mu \bar{\psi}^i u_i^{-}
        + 4 i \lambda^{(1,0)}{}^{\alpha} 
                        ( \varepsilon C^{++} \sigma^\mu \bar{\psi}^i )_\alpha u_i^{-} 
        \nonumber\\
&&{} 
        - \partial^{++} \lambda_\mu^{(1,1)} 
        - \sqrt{2} C^{++}{}^{\alpha\beta} 
                        ( \sigma_\mu \bar{\sigma}^\nu \varepsilon)_{\alpha\beta}
                        \lambda_\nu^{(1,1)} \bar{\phi}
        , 
        \label{eq:EqsToDetermineDeformedSUSY(1,1)}
        \\
4 
\delta^{*}_{\xi} \psi_{\alpha}^i u_i^{-} 
&=& 
        - 2 (\sigma^\mu \bar{\sigma}^\nu \xi^{-})_{\alpha} 
                        \partial_\nu A_\mu 
        + 6 \xi^{+}_{\alpha} D^{ij} u_i^{-} u_j^{-} 
                        \nonumber\\
&&{} 
        - i (\sigma^\nu \partial_\nu \lambda^{(0,1)})_\alpha 
        - 2 \sqrt{2} i ( \varepsilon C^{+-} 
                        \sigma^\nu \partial_\nu \lambda^{(0,1)} )_\alpha \bar{\phi} 
                        \nonumber\\
&&{} 
        - 6 i (\varepsilon C^{++} \lambda^{(1,0)})_{\alpha} D^{ij} u_i^{-} u_j^{-} 
        + 2 i \lambda^{(1,0)}{}^{\beta} 
                        (\varepsilon C^{+-} \sigma^\nu 
                        \bar{\sigma}^\mu \varepsilon)_{\beta\alpha} 
                        \partial_\nu A_\mu 
                        \nonumber\\
&&{} 
        + 2 i \partial_\nu \lambda^{(1,0)}_{\alpha} C^{+-}{}^{\beta\gamma} 
                        (\sigma^\mu \bar{\sigma}^\nu \varepsilon)_{\beta\gamma} 
                        A_\mu 
        - 2 i (\sigma^\mu \bar{\psi}^i )_{\alpha} u_i^{-} 
                        C^{++}{}^{\gamma\delta} 
                        (\sigma_\mu \bar{\sigma}^\nu \varepsilon)_{\gamma\delta}
                        \lambda_\nu^{(1,1)} 
                        \nonumber\\
&&{} 
        - \partial^{++} \lambda^{(1,2)}_\alpha 
        - 2 \sqrt{2} ( \varepsilon C^{++} \lambda^{(1,2)} )_\alpha \bar{\phi} 
        + (\varepsilon C^{++} \sigma^\mu \lambda^{(2,1)} )_\alpha A_\mu 
        , 
        \label{eq:EqsToDetermineDeformedSUSY(1,2)}
        \\
-4 
\delta^{*}_{\xi} \bar{\psi}^{\dot{\alpha} i} u_i^{-}
&=& 
        2 \sqrt{2} (\xi^{-} \sigma^\mu)_{\dot{\beta}} 
                        \varepsilon^{\dot{\beta}\dot{\alpha}} 
                        \partial_\mu \bar{\phi} 
        + i \partial_\mu \lambda^{(1,0)}{}^\alpha 
                        \sigma^\mu_{\alpha{\dot{\beta}}} 
                        \varepsilon^{\dot{\beta}\dot{\alpha}} 
                        \nonumber\\
&&{} 
        + 2 \sqrt{2} i \partial_\nu 
                \left\{ 
                \lambda^{(1,0)}{}^{\alpha} 
                        (\varepsilon C^{+-} \sigma^\nu )_{\alpha{\dot{\beta}}} 
                        \varepsilon^{{\dot{\beta}}{\dot{\alpha}}} 
                        \bar{\phi} 
                \right\} 
        - \partial^{++} \lambda^{(2,1)}{}^{\dot{\alpha}} 
        , 
        \label{eq:EqsToDetermineDeformedSUSY(2,1)}
        \\
3
\delta^{*}_{\xi} D^{ij} u_i^{-} u_j^{-} 
&=& 
        - 4 i \xi^{-} \sigma^\mu \partial_\mu \bar{\psi}^i u_i^{-}
        + 4 \partial_\nu 
                \left\{ 
                \lambda^{(1,0)}{}^{\alpha} 
                        (\varepsilon C^{+-} \sigma^\nu \bar{\psi}^i )_{\alpha} u_i^{-} 
                \right\} 
                        \nonumber\\
&&{} 
        - i \partial^\mu \lambda_\mu^{(1,1)} 
        - \sqrt{2} i C^{+-}{}^{\alpha\beta} 
                        ( \sigma^\mu \bar{\sigma}^\nu \varepsilon )_{\alpha\beta} 
                        \partial_\nu ( \lambda_\mu^{(1,1)} \bar{\phi} )
        - \partial^{++} \lambda^{(2,2)}
. 
\label{eq:EqsToDetermineDeformedSUSY(2,2)}
\end{eqnarray} 
Inserting the harmonic expansions of gauge parameters into the above
equations,  one obtains a set of 
recursive relations for harmonic modes, which can be solved order by order
in $C$.
Up to the $O(C)$ terms, 
the associated gauge parameter $\Lambda$ is given by the following components: 
\begin{eqnarray}
\lambda^{(1,0)}{}^{\alpha}  
&=&
        -i2\sqrt{2} \xi^-{}^{\alpha} \bar{\phi} 
        +i4 (\xi_m \varepsilon C_{(kl)})^{\alpha} \bar{\phi}^2  
        \left( 
          u^+{}^{(k} u^-{}^l u^-{}^{m)} 
         -\frac{8}{3} \epsilon^{k(l} u^-{}^{m)}
        \right)
        + O(C^2) 
        , \nonumber\\ 
\lambda^{(0,1)}_{\dot{\alpha}}  
&=&
        -i2 (\xi^- \sigma^{\mu})_{\dot{\alpha}} A_{\mu} 
                \nonumber\\
&&{} 
        +i2\sqrt{2} 
                (\xi_m \varepsilon C_{(kl)} \sigma^{\mu})_{\dot{\alpha}} 
                \bar{\phi} A_{\mu}  
                \left( 
                  u^+{}^{(k} u^-{}^l u^-{}^{m)} 
                 -\frac{8}{3} \epsilon^{k(l} u^-{}^{m)}
                \right)
        + O(C^2) 
        , \nonumber\\ 
\lambda^{(1,1)}_{\mu}  
&=& 
        2(\xi^- \sigma_{\mu} \bar{\psi}^-) 
                \nonumber\\
&&{} 
        +4\sqrt{2} (\xi_m \varepsilon C_{(kl)} \sigma_{\mu} \bar{\psi}_n) 
                \bar{\phi} 
                \left( 
                u^+{}^{(k} u^-{}^l u^-{}^m u^-{}^{n)} 
                 -\frac{9}{4} \epsilon^{k(l} u^-{}^m u^-{}^{n)} 
                \right)
        + O(C^2) 
        , \nonumber\\ 
\lambda^{(0,2)}  
&=& 
        2(\xi^- \psi^-) 
        +\frac{8\sqrt{2}}{3} (\xi_m \varepsilon C_{(kl)} \psi_n) \bar{\phi}  
                \left( 
                u^+{}^{(k} u^-{}^l u^-{}^m u^-{}^{n)} 
                 -\frac{9}{4} \epsilon^{k(l} u^-{}^m u^-{}^{n)} 
                \right)
        \nonumber\\ 
&&{} 
        +\frac{4}{3} (\xi_m \varepsilon C_{(kl)} \sigma^{\mu} \bar{\psi}_n) 
                A_{\mu}  
                \left( 
                u^+{}^{(k} u^-{}^l u^-{}^m u^-{}^{n)} 
                 -\frac{9}{4} \epsilon^{k(l} u^-{}^m u^-{}^{n)} 
                \right)
        + O(C^2) 
        , \nonumber\\ 
\lambda^{(1,2)}{}^{\alpha}  
&=& 
        2\xi^-{}^{\alpha} D^{--}
        , \nonumber\\  
&&{} 
        -4\sqrt{2} (\xi_p \varepsilon C_{(kl)})^{\alpha} D_{mn} 
                \bar{\phi}  
                \left( 
                u^+{}^{(k} u^-{}^l u^-{}^m u^-{}^n u^-{}^{p)} 
                 -\frac{32}{15} 
                \epsilon^{k(l} u^-{}^m u^-{}^n u^-{}^{p)} 
                \right)
                \nonumber\\ 
&&{} 
        + i 4 (\xi_p \varepsilon C_{(kl)})^{\alpha} 
                (\bar{\psi}_m \bar{\psi}_n)  
                \left( 
                u^+{}^{(k} u^-{}^l u^-{}^m u^-{}^n u^-{}^{p)} 
                 -\frac{32}{15} 
                \epsilon^{k(l} u^-{}^m u^-{}^n u^-{}^{p)} 
                \right)
                \nonumber\\ 
&&{} 
        + 4\sqrt{2} (\xi_m \varepsilon C_{(kl)} \sigma^{\mu\nu})^{\alpha} 
                \partial_{\mu} (A_{\nu} \bar{\phi})  
                u^-{}^{(k} u^-{}^l u^-{}^{m)}
                \nonumber\\ 
&&{} 
        - \frac{8\sqrt{2}}{3} 
                (\xi_m \sigma^{\mu\nu}\varepsilon C_{(kl)})^{\alpha} 
                ( \partial_{\mu} A_{\nu} \bar{\phi} 
        -A_{\nu} \partial_{\mu} \bar{\phi} )  
                u^-{}^{(k} u^-{}^l u^-{}^{m)}
                \nonumber\\ 
&&{} 
        + \frac{2\sqrt{2}}{3} (\xi_m \varepsilon C_{(kl)})^{\alpha} 
                A_{\mu} \partial^{\mu} \bar{\phi}  
                u^-{}^{(k} u^-{}^l u^-{}^{m)}
                \nonumber\\ 
&&{} 
        -2\sqrt{2} (\xi_m \varepsilon C_{(kl)})^{\alpha} 
                \partial^{\mu} A_{\mu} \bar{\phi}  
                u^-{}^{(k} u^-{}^l u^-{}^{m)} 
        + O(C^2) 
        , \nonumber\\
\lambda^{(2,1)}_{\dot{\alpha}} 
&=&
        +4 (\xi_m \varepsilon C_{(kl)}\sigma^{\mu})_{\dot{\alpha}} 
                \partial_{\mu} (\bar{\phi})^2 
                u^-{}^{(k} u^-{}^l u^-{}^{m)} 
        + O(C^2) 
        ,\nonumber\\ 
\lambda^{(2,2)} 
&=& 
        -i4\sqrt{2} (\xi_m \varepsilon C_{(kl)} \sigma^{\mu})_{\dot{\alpha}} 
                \partial_{\mu} (\bar{\psi}_n^{\dot{\alpha}} \bar{\phi}) 
                u^-{}^{(k} u^-{}^l u^-{}^m u^-{}^{n)} 
        + O(C^2) 
. 
\end{eqnarray}
Then we find 
the deformed supersymmetry transformation laws in the WZ gauge: 
\begin{eqnarray}
\delta^{*}_{\xi} \phi 
&=& 
        - \sqrt{2} i \xi^{i} \psi_{i}
        - \frac83 i ( \xi^{j} \varepsilon C_{(j k)} \psi^{k} ) \bar{\phi} 
        - {2 \sqrt{2} \over 3} i 
                ( \xi^{j} \varepsilon C_{(j k)} \sigma^\nu \bar{\psi}^{k} ) 
                A_\nu
        + O(C^2)
        , \nonumber \\
\delta^{*}_{\xi} \bar{\phi}
&=& 
        0 
        , \nonumber \\
\delta^{*}_{\xi} A_\mu 
&=& 
        i \xi^{i} \sigma_\mu \bar{\psi}_{i}
        + 2 \sqrt{2}i ( \xi^{j} \varepsilon C_{(j k)} \sigma_\mu \bar{\psi}^{k} ) 
                \bar{\phi}
        + O(C^2)
        , \nonumber \\
\delta^{*}_{\xi} \psi^{\alpha i} 
&=& 
        - (\xi^{i} \sigma^{\mu\nu} )^{\alpha} F_{\mu\nu}         
        - D^{ij} \xi^\alpha_j 
        + 2 \sqrt{2} D^{(i j} ( \xi^{k)} \varepsilon C_{(j k)} )^\alpha 
                \bar{\phi} 
        - 2i (\bar{\psi}^{(i} \bar{\psi}^{j}) 
                ( \xi^{k)} \varepsilon C_{(j k)} )^\alpha  
                        \nonumber\\
&&{} 
        - \left\{  
                2 \sqrt{2} (\xi^{j} \varepsilon C_{(j k)} 
                        \sigma^{\mu\nu} )^\alpha 
                + {2\sqrt{2}\over 3} (\xi^{j} \sigma^{\mu\nu} \varepsilon 
                        C_{(j k)} )^\alpha 
                + \sqrt{2} C_{(j k)}^{\mu\nu} \xi^{\alpha j} 
        \right\} 
                \epsilon^{k i} \bar{\phi} F_{\mu\nu} 
                        \nonumber\\
&&{} 
        + \left\{ 
                {4\sqrt{2} \over 3} (\xi^{j} 
                        \sigma^{\mu\nu} \varepsilon C_{(j k)} )^\alpha 
                + 2 \sqrt{2} C_{(j k)}^{\mu\nu} \xi^{\alpha j} 
        \right\} 
                \epsilon^{k i} \partial_\mu \bar{\phi} A_\nu 
                        \nonumber\\
&&{} 
        - {2\sqrt{2} \over 3} 
                (\xi^{j} \varepsilon C_{(j k)} )^\alpha 
                \epsilon^{k i} \partial^\mu \bar{\phi} A_\mu 
        + O(C^2)
        , \nonumber \\
\delta^{*}_{\xi} \bar{\psi}^{i}_{\dot{\alpha}}  
&=& 
        + \sqrt{2} (\xi^{i} \sigma^\nu )_{\dot{\alpha}} 
                \partial_\nu \bar{\phi} 
        + 2 
                (\xi^{j} \varepsilon C_{(j k)} \sigma^\nu)_{\dot{\alpha}} 
                \partial_\nu (\bar{\phi}^2) 
                \epsilon^{k i}
        + O(C^2) 
        , \nonumber \\
\delta^{*}_{\xi} D^{ij} 
&=& 
        - 2 i \xi^{(i} \sigma^\nu \partial_\nu \bar{\psi}^{j)}
        - 6 \sqrt{2} i 
                \epsilon^{k(l} \partial_\nu \bigl\{
                (\xi^{i} \varepsilon C_{(k l)} \sigma^\nu 
                \bar{\psi}^{j)}) \bar{\phi} \bigr\} 
        + O(C^2)
. 
\label{eq:DeformedSUSYTr1}
\end{eqnarray} 
To obtain the expression for $\delta^{*}_{\xi} \psi$, 
we have used the following relation: 
\begin{equation} 
(\xi^{i} \varepsilon C_{(j k)} \sigma^{\mu\nu} )^\alpha 
+ (\xi^{i} \sigma^{\mu\nu} \varepsilon C_{(j k)} )^\alpha 
+ \xi^{\alpha i} C_{(j k)}^{\mu\nu}
= 
        0 
        ,
\label{eq:Relation} 
\end{equation} 
which can be proved by explicit calculation.
Note that the expression for $\delta^{*}_{\xi} \psi$ given above 
is one of the possible expressions and is chosen so that 
the invariance of the action can be easily examined.

We can check that the $O(C)$ action (\ref{eq:O(C)action}) is 
indeed invariant under the deformed supersymmetry transformation 
(\ref{eq:DeformedSUSYTr1}).
Denoting the deformed supersymmetry transformation $\delta^{*}_{\xi}$ as 
\begin{equation} 
\delta^{*}_{\xi} 
= 
        \delta^{*}_{\xi}{}^{(0)} 
        + \delta^{*}_{\xi}{}^{(1)}  
        + \cdots 
        , 
\label{eq:ExpandedSUSYTr} 
\end{equation} 
where $\delta^{*}_{\xi}{}^{(n)}$ represents the $O(C^n)$ variations, 
we can see that 
\begin{equation} 
\delta^{*}_{\xi}{}^{(1)} S_{*,2} + \delta^{*}_{\xi}{}^{(0)} S_{*,3} 
= 
        0. 
\label{eq:Cancellation} 
\end{equation} 
To show this, we need (\ref{eq:Relation}) 
and a formula 
$
T_{ij} - T_{ji} 
= 
	\epsilon_{ij} \epsilon^{kl} T_{lk} 
$ 
for a given tensor $T_{ij}$. 
Note that $\delta^{*}_{\xi}{}^{(1)}S_{*,3}$ is non zero for generic
deformation parameters. 
Therefore we need higher order terms in $C$ in order to obtain fully 
supersymmetric action.

After the redefinition (\ref{eq:orderCgaugetr}), 
the deformed supersymmetry transformation (\ref{eq:DeformedSUSYTr1}) becomes 
\begin{eqnarray}
\delta^{*}_{\xi} \hat{\phi} 
&=& 
        - \sqrt{2} i \xi^{i} \hat{\psi}_{i}
        - \frac83 i ( \xi^{j} \varepsilon C_{(j k)} \hat{\psi}^{k} ) \hat{\bar{\phi}} 
        + O(C^2)
        , \nonumber \\
\delta^{*}_{\xi} \hat{\bar{\phi}}
&=& 
        0 
        , \nonumber \\
\delta^{*}_{\xi} \hat{A}_\mu 
&=& 
        i \xi^{i} \sigma_\mu \hat{\bar{\psi}}{}_{i}
        + 2 \sqrt{2}i ( \xi^{j} \varepsilon C_{(j k)} \sigma_\mu \hat{\bar{\psi}}{}^{k} ) 
                \hat{\bar{\phi}}
        + O(C^2)
        , \nonumber \\
\delta^{*}_{\xi} \hat{\psi}^{\alpha i} 
&=& 
        - (\xi^{i} \sigma^{\mu\nu} )^{\alpha} \hat{F}_{\mu\nu}   
        - \hat{D}^{ij} \xi^\alpha_j 
        - i ( \xi^{i} \sigma_{\mu\nu} )^\alpha 
                C_{(j k)}^{\mu\nu} (\hat{\bar{\psi}}{}^{j} \hat{\bar{\psi}}{}^{k}) 
        + 2 \sqrt{2} \hat{D}^{(i j} ( \xi^{k)} \varepsilon C_{(j k)} )^\alpha 
                \hat{\bar{\phi}} 
                        \nonumber\\
&&{} 
        - \left\{  
                2 \sqrt{2} (\xi^{j} \varepsilon C_{(j k)} 
                        \sigma^{\mu\nu} )^\alpha 
                + {2\sqrt{2}\over 3} (\xi^{j} \sigma^{\mu\nu} \varepsilon 
                        C_{(j k)} )^\alpha 
                + \sqrt{2} C_{(j k)}^{\mu\nu} \xi^{\alpha j} 
        \right\} 
                \epsilon^{k i} \hat{\bar{\phi}} \hat{F}_{\mu\nu} 
        + O(C^2)
        , \nonumber \\ 
\delta^{*}_{\xi} \hat{\bar{\psi}}{}^{i}_{\dot{\alpha}}  
&=& 
        + \sqrt{2} (\xi^{i} \sigma^\nu )_{\dot{\alpha}} 
                \partial_\nu \hat{\bar{\phi}} 
        + 2 
                (\xi^{j} \varepsilon C_{(j k)} \sigma^\nu)_{\dot{\alpha}} 
                \partial_\nu (\hat{\bar{\phi}}{}^2) 
                \epsilon^{k i}
        + O(C^2) 
        , \nonumber \\
\delta^{*}_{\xi} \hat{D}^{ij} 
&=& 
        - 2 i \xi^{(i} \sigma^\nu \partial_\nu \hat{\bar{\psi}}{}^{j)}
                        \nonumber\\
&&{} 
        - 6 \sqrt{2} i 
                \epsilon^{k(l} \partial_\nu \bigl\{
                (\xi^{i} \varepsilon C_{(k l)} \sigma^\nu 
                \hat{\bar{\psi}}{}^{j)}) \hat{\bar{\phi}} \bigr\} 
        + 2\sqrt{2} i \epsilon^{il}\epsilon^{jm} 
			( \xi^k \varepsilon C_{(lm)} \sigma^\nu \hat{\bar{\psi}}{}_k ) 
                \partial_\nu \hat{\bar{\phi}} 
        + O(C^2)
. 
\label{eq:DeformedSUSYTr2}
\end{eqnarray} 
For generic non-singlet deformations, it seems difficult to find the
appropriate field redefinition such that both gauge and supersymmetry
transformations become canonical. 

In this paper we have studied ${\cal N}=2$ supersymmetric $U(1)$ gauge theory
in non(anti)-commutative harmonic superspace with the non-singlet deformation
parameter $C$. 
We have determined 
deformed ${\cal N}=(1,0)$ supersymmetry transformation at the order $C$
for component
fields of the analytic superfield $V^{++}_{WZ}$ in the WZ gauge.
We have checked that the $O(C)$ component action is invariant under this 
deformed supersymmetry transformation.

It is interesting to study
 the reduction of deformation parameters such that only
${\cal N}=1$ subspace becomes non(anti)commutative.
In this case we will be able to construct gauge and ${\cal N}=(1,0)$
supersymmetry transformations.
The action  
(\ref{eq:StarDeformedAction:Gen}) will reduce to the component action
defined in 
${\cal N}=1/2$ superspace by some field identifications,
which is expected to have ${\cal N}=(1,1/2)$ supersymmetry\cite{IvLeZu}.
A detailed analysis will appear in a forthcoming paper\cite{ArItOh4}.

Another obvious generalization is the extension to non-abelian gauge groups. 
For a  gauge group $U(N)$, it would be possible to construct the ${\cal
N}=(1,0)$ supersymmetry in a similar way.
In particular, it would be interesting to study the (deformed) 
central charge in the algebra.
Instanton solutions in the deformed gauge theory and its
contribution to the prepotential of the low-energy effective theory will be
also interesting in viewpoint of its relation to superstring theory with R-R
background.

{\bf Acknowledgments}:
T.~A. is supported by 
the Grant-in-Aid for Scientific Research in Priority Areas (No.14046201) 
from the Ministry of Education, Culture, Sports, Science 
and Technology. 
A.~O. is supported by a 21st Century COE Program at 
Tokyo Tech "Nanometer-Scale Quantum Physics" by the 
Ministry of Education, Culture, Sports, Science and Technology.


\begin{thebibliography}{99}

\bibitem{ncsuper}
J.~H.~Schwarz and P.~Van Nieuwenhuizen,
Lett.\ Nuovo Cim.\  {\bf 34}, 21 (1982). 

 \bibitem{OoVa} H.~Ooguri and C.~Vafa,
Adv. Theor. Math. Phys. {\bf 7} (2003) 53,
hep-th/0302109; 
Adv. Theor. Math. Phys. {\bf 7} (2004) 405, hep-th/0303063. 

\bibitem{BeSe}
N.~Berkovits and N.~Seiberg,
JHEP {\bf 0307} (2003) 010,
hep-th/0306226. 


 \bibitem{DeGrNi} J.~de Boer, P.~A.~Grassi and P.~van Nieuwenhuizen,
Phys. Lett. {\bf B574} (2003) 98,
hep-th/0302078. 

 \bibitem{Se} N.~Seiberg,
JHEP {\bf 0306}, 010 (2003),
hep-th/0305248. 

\bibitem{ArItOh1}
T.~Araki, K.~Ito and A.~Ohtsuka, 
Phys. Lett. {\bf B573} (2003) 209, 
hep-th/0307076. 

\bibitem{Pert}
R.~Britto, B.~Feng and S.~J.~Rey,
JHEP {\bf 0307} (2003) 067,
hep-th/0306215; \\
S.~Terashima and J.~T.~Yee,
JHEP {\bf 0312} (2003) 053, hep-th/0306237; \\
R.~Britto, B.~Feng and S.~J.~Rey,
JHEP {\bf 0308} (2003) 001,
hep-th/0307091;\\
M.~T.~Grisaru, S.~Penati and A.~Romagnoni, 
JHEP {\bf 0308} (2003) 003, 
hep-th/0307099;\\
R.~Britto and B.~Feng, 
Phys. Rev. Lett. {\bf 91} (2003) 201601,
hep-th/0307165;\\
A.~Romagnoni, 
JHEP {\bf 0310} (2003) 016, 
hep-th/0307209;\\
O.~Lunin and  S.~J.~Rey,
 JHEP {\bf 0309} (2003) 045,
hep-th/0307275;\\
D.~Berenstein and S.~J.~Rey, 
Phys. Rev. {\bf D68} (2003) 121701,
hep-th/0308049;\\
M.~Alishahiha, A.~Ghodsi and N.~Sadooghi, 
Nucl.\ Phys.\ B {\bf 691} (2004) 111,
hep-th/0309037;\\
A.~T.~Banin, I.~L.~Buchbinder and N.~G.~Pletnev,
JHEP {\bf 0407} (2004) 011
hep-th/0405063;\\
A.~Sako and T.~Suzuki,
hep-th/0408226.

\bibitem{Inst}
A.~Imaanpur, 
JHEP {\bf 0309} (2003) 077, hep-th/0308171;
JHEP {\bf 0312} (2003) 009, 
hep-th/0311137;\\
P.~A.~Grassi, R.~Ricci and D.~Robles-Llana, 
JHEP {\bf 0407} (2004) 065,
hep-th/0311155;\\
R.~Britto, B.~Feng, O.~Lunin and S.~J.~Rey, 
Phys.\ Rev.\ D {\bf 69} (2004) 126004,
hep-th/0311275;\\
M.~Billo, M.~Frau, I.~Pesando and A.~Lerda,
JHEP {\bf 0405} (2004) 023,
arXive:hep-th/0402160;\\ 
A.~Imaanpur and S.~Parvizi,
JHEP {\bf 0407} (2004) 010, 
hep-th/0403174.


\bibitem{KlPeTa}
D.~Klemm, S.~Penati and L.~Tamassia,
Class.\ Quant.\ Grav.\  {\bf 20} (2003) 2905,
hep-th/0104190; \\ 
S.~Ferrara and M.~A.~Lledo,
JHEP {\bf 0005} (2000) 008,
hep-th/0002084; \\
S.~Ferrara, M.~A.~Lledo and O.~Macia,
JHEP {\bf 0309} (2003) 068,
hep-th/0307039.  

\bibitem{HaIsUm}
M.~Hatsuda, S.~Iso and H.~Umetsu,
Nucl. Phys. {\bf B671} (2003) 217,
hep-th/0306251 


\bibitem{IvLeZu}
E. Ivanov, O. Lechtenfeld and B. Zupnik,
JHEP {\bf 0402} (2004) 012,
hep-th/0308012;
hep-th/0408146.

\bibitem{FeSo}
S.~Ferrara and E.~Sokatchev, 
Phys. Lett. {\bf B579} (2004) 226,
hep-th/0308021.


\bibitem{ArItOh2}
T.~Araki, K.~Ito and A.~Ohtsuka, 
JHEP {\bf 0401} (2004) 046, 
hep-th/0401012.

\bibitem{SaWo}
C.~S\"amann and M.~Wolf, 
JHEP {\bf 0403} (2004) 048,
hep-th/0401147.

\bibitem{KeSa}
S.V.~Ketov and S.~Sasaki,
Phys.\ Lett.\ B {\bf 595} (2004) 530,
hep-th/0404119;
Phys.\ Lett.\ B {\bf 597} (2004) 105,
hep-th/0405278;
hep-th/0407211.

\bibitem{ArIt3}
T.~Araki and K.~Ito, 
Phys. Lett. {\bf B595} (2004) 513, 
hep-th/0404250.


\bibitem{FeIvLeSoZu}
S.~Ferrara, E.~Ivanov, O.~Lechtenfeld, E.~Sokatchev and B.~Zupnik,
hep-th/0405049.

\bibitem{SeWi} N.~Seiberg and E.~Witten, 
JHEP {\bf 9909} (1999) 032,
hep-th/9908142.

\bibitem{Mi}
D.~Mikulovic, 
JHEP {\bf 0401} (2004) 063,
hep-th/0310065;
JHEP {\bf 0405} (2004) 077,
hep-th/0403290.

\bibitem{WeBa}
J.~Wess and J.~Bagger, ``Supersymmetry and
Supergravity,''
Princeton University Press, 1992.


\bibitem{GaIvOgSo}
A.~Galperin, E.~Ivanov, V.~Ogievetsky and E.~Sokatchev,
``Harmonic Superspace'', Cambridge University Press , 2001.

\bibitem{Zu}
 B.M.~Zupnik, 
Phys. Lett. {\bf B183} (1987) 175.

\bibitem{BuSa}
I.~L.~Buchbinder and I.~B.~Samsonov,
Grav.Cosmol.{\bf 8} (2002) 17, 
hep-th/0109130.

\bibitem{ArItOh4}
T.~Araki, K.~Ito and A.~Ohtsuka, work in progress.
\end{thebibliography}
\end{document}